\documentclass[12pt]{iopart}

\pdfoutput=1

\usepackage{mathptmx}
\usepackage[scaled=0.9]{helvet}
\usepackage[utf8]{inputenc} 
\usepackage{graphicx}
\usepackage[dvipsnames]{xcolor}
\usepackage{epstopdf}
\usepackage[textwidth=1.2cm]{todonotes}
\usepackage{textcomp}
\usepackage{sidecap}
\usepackage[
,textwidth=17.5cm
,textheight=23.5cm
,verbose
,dvips
]{geometry}
\usepackage{xspace}
\usepackage{lipsum}
\usepackage{gensymb}
\usepackage{tabularx}
\usepackage{iopams}
\usepackage{cite}

\newcommand{\doxB}{$(D^{0},X_\textrm{B})$}
\newcommand{\SioxA}{$(\textrm{Si}^{0},X_\textrm{A})$}
\newcommand{\OoxA}{$(\textrm{O}^{0},X_\textrm{A})$}
\newcommand{\aoxA}{$(A^{0},X_\textrm{A})$}
\newcommand{\uxA}{$UX1$}
\newcommand{\uxB}{$UX2$}

\newcommand{\xA}{$X_\textrm{A}$}

\begin{document}

\title[Improved control over spontaneously formed GaN NWs in MBE using a two-step growth process]{Improved control over spontaneously formed GaN nanowires in molecular beam epitaxy using a two-step growth process}

\author{J K Zettler, P Corfdir, L Geelhaar, H Riechert, O Brandt and S Fernández-Garrido}
\address{Paul-Drude-Institut für Festkörperelektronik, Hausvogteiplatz 5–7, 10117 Berlin, Germany}
\ead{garrido@pdi-berlin.de}


\begin{abstract}
We investigate the influence of modified growth conditions during the spontaneous formation of GaN nanowires on Si$(111)$ in plasma-assisted molecular beam epitaxy. We find that a two-step growth approach, where the substrate temperature is increased during the nucleation stage, is an efficient method to gain control over the area coverage, average diameter, and coalescence degree of GaN nanowire ensembles. Furthermore, we also demonstrate that the growth conditions employed during the incubation time that precedes nanowire nucleation do not influence the properties of the final nanowire ensemble. Therefore, when growing GaN nanowires at elevated temperatures or with low Ga/N ratios, the total growth time can be reduced significantly by using more favorable growth conditions for nanowire nucleation during the incubation time.
\end{abstract}

\maketitle

\section{Introduction}

Unlike GaN films, single-crystalline GaN nanowires (NWs) can be grown on a wide variety of crystalline as well as amorphous substrates by several epitaxial growth techniques~\cite{Li2012,Chen2010a,Yoshizawa1997,Sanchez-Garcia1998,Calleja2007,Calarco2007,Stoica2008,Bertness2011,Geelhaar_ieeejstqe_2011,Schuster2012,Sobanska2014}. In plasma-assisted molecular beam epitaxy (PA-MBE), dense ensembles of GaN NWs form spontaneously under N-excess at elevated temperatures~\cite{Fernandez-Garrido_jap_2009,Bertness2011,Consonni2013,Fernandez-Garrido2013a,Geelhaar_ieeejstqe_2011}. Regardless of the substrate, spontaneously formed single GaN NWs are virtually free of both extended defects and homogeneous strain~\cite{Calleja2000,Trampert2003}. Nevertheless, as for any other self-organized growth process, the degree of control on the properties of GaN NW ensembles is rather limited.

In PA-MBE, the diameter of GaN NWs can be regulated by the Ga/N ratio~\cite{Bertness2011,Fernandez-Garrido2013a} and the NW number density is expected to be determined by the diffusion length of Ga adatoms~\cite{Ristic2008}. For low Ga/N ratios, the diameter of single NWs can be as small as $10-15$~nm~\cite{Consonni2011,Fernandez-Garrido_jap_2009,Calarco2007,Stoica2008}. However, in most cases, the effective average diameter becomes larger due to the coalescence of adjacent NWs. This undesired phenomenon, caused by NW mutual misorientation as well as NW radial growth~\cite{Brandt2014}, is favored by the high nucleation density of GaN NWs (typically, in the range of $10^{9}-10^{10}$~cm$^{-2}$)~\cite{Calarco2007,Fernandez-Garrido_jap_2009,Consonni2011a}. Unfortunately, NW coalescence not only results in a poor degree of control over the NW morphology but also introduces extended defects as well as inhomogeneous strain~\cite{Consonni2009,Jenichen2011a,Kaganer2012,Grossklaus2013a,Fernandez-Garrido2014}. Therefore, in order to take advantage of spontaneously formed NWs for the fabrication of GaN-based devices on dissimilar substrates, it is highly desirable to develop growth approaches designed to gain control over the nucleation and growth of GaN NWs.

Carnevale et al.~\cite{Carnevale2011} proposed a two-step growth method to separate the nucleation and growth processes. They showed that, by increasing the substrate temperature during the nucleation stage, it is possible to disentangle and vary the number density, the height, and the length of spontaneously formed GaN NWs. However, even though they were able to vary the NW number density in a wide range, the resulting NW ensembles were rather short ($<400~$nm) and inhomogeneous in height. 

Motivated by the experiments performed by Carnevale et al.~\cite{Carnevale2011}, we systematically investigate in the present work the impact of modifying the growth conditions at different stages during the spontaneous formation of homogeneous GaN NW ensembles on Si. In the first part of this work, we analyze the impact of increasing the substrate temperature at different times during the nucleation of GaN NWs. The latter is monitored in situ by reflection high-energy electron diffraction (RHEED) as well as line-of-sight quadrupole mass spectrometry (QMS)~\cite{Consonni2011,Consonni_apl_2011b,Cheze_apl_2010a,Limbach2012,Fernandez-Garrido2015}. While we were not able to reduce significantly the number density for ensembles of long and homogeneous NWs, we gained control over area coverage, average NW diameter and coalescence degree. In the second part of this work, we investigate the influence of the growth conditions employed during the incubation stage~\cite{Calarco2007,Consonni_apl_2011b,Fernandez-Garrido2015} that precedes NW nucleation. The results demonstrate that the properties of the NW ensemble do not depend on the incubation stage. Therefore, a two-step growth approach, where more favorable growth conditions for NW nucleation are employed during the incubation stage, can be used to reduce the total growth time without affecting the properties of the final NW ensemble. This growth approach paves the way for growing GaN NWs under more extreme growth conditions (lower Ga/N ratios and higher substrate temperatures) which typically result in smaller NW diameters, lower degrees of coalescence, and improved optical properties~\cite{Corfdir2014,zettler2015,Lefebvre2011,Bertness2011,Fernandez-Garrido_jap_2009,Geelhaar_ieeejstqe_2011}.

\section{Experiments and methods}

All samples were grown on $2^{\prime\prime}$ Si$(111)$ substrates in a MBE system equipped with a solid-source effusion cell for Ga as well as a radio-frequency N$_{2}$ plasma source for active N. The impinging fluxes were calibrated in GaN-equivalent growth rate units of nm/min, as described in Ref.~\cite{Heying2000}. A growth rate of 1~nm/min is equivalent to $7.3\times 10^{13}$~\textrm{atoms}~cm$^{-2}$~s$^{-1}$. The desorbing Ga flux $\Phi_\textrm{des}$ was monitored in-situ during the experiments by QMS. The QMS response to the Ga$^{69}$ signal was also calibrated in GaN-equivalent growth rate units, as explained in detail elsewhere~\cite{Koblmuller2004,Brown2006,Fernandez-Garrido2015}. Since there is no Ga accumulation on the surface~\cite{Ristic2008,Fernandez-Garrido2013a,Consonni2013}, the Ga incorporation rate per unit area $\Phi_\textrm{inc}$ (i.\,e., the deposition rate) is given by $\Phi_\textrm{Ga} - \Phi_\textrm{des}$, where $\Phi_\textrm{Ga}$ is the impinging Ga flux. The substrate temperature was measured with an optical pyrometer calibrated to the $1\times1\rightarrow7\times7$ surface reconstruction transition temperature of Si$(111)$ (approx. $860^{\circ}$C).~\cite{Suzuki1993} The as-received Si substrates were etched using diluted ($5\%$) HF. In order to remove any residual Si$_{x}$O$_{y}$ from the surface, the substrates were outgassed at $885^{\circ}$C for $30$~min prior to growth. Afterward, the substrates were exposed to an active nitrogen flux of $\Phi_\textrm{N} = (11.0 \pm 0.5)$~nm/min at the growth temperature for $10$~min. Then, the growth was initiated by opening the Ga shutter at $t = 0$. For all experiments $\Phi_\textrm{N}$ was kept constant and equal to $(11.0 \pm 0.5)$~nm/min. For the two-step growth samples, the substrate temperature was first kept at $815$\celsius. Then, the substrate temperature was increased to $845$\celsius\xspace at the specified times during the NW nucleation stage at a rate of $10$\celsius /min keeping the Ga and N shutters open during the entire process. For the two-step sample in section \ref{sec2}, the substrate temperature was instead increased to $855$\celsius\xspace and subsequently the Ga flux was raised as well. At the end, the growth was stopped by closing all shutters and cooling down the samples.

The morphology of the samples was investigated by cross-sectional and plan-view scanning electron microscopy following the method established in Ref.~\cite{Brandt2014}. For each sample, we analyzed several plan-view micrographs containing a few hundreds of NWs. Using the open-source software ImageJ~\cite{Schneider2012a}, we derived the area $A$, perimeter $P$, and circularity $C$ of the NW top facets. The circularity is defined by
\begin{equation}
\label{equation1}
C=\frac{4\pi{A}}{P^{2}}
\end{equation}
and was used to estimate the coalescence degree $\sigma_{C}$ as explained below.

While uncoalesced NWs typically exhibit high values of $C$, anisotropic radial growth as well as NW coalescence usually lead to shapes exhibiting lower values of $C$~\cite{Brandt2014}. As proposed in Ref.~\cite{Brandt2014}, we used a threshold value of $C < \zeta_{A} = 0.762$ to distinguish between single NWs and coalesced aggregates. Furthermore, we introduced an additional criterion considering only NWs that also exhibit an equivalent-disk diameter $d < 100$~nm as uncoalesced. This is due to the fact that for highly coalesced NW ensembles NW aggregates may cluster as bundles with roundish cross-sectional shapes exhibiting circularity values higher than $\zeta_{A}$. The coalescence degree was then assessed as 
\begin{equation}
\label{equation2}
\sigma_{C}=\frac{A_{C}}{A_{T}},
\end{equation}
where $A_{C}$ is the total cross-sectional area of coalesced NWs and ${A_{T}}$ the total cross-sectional area of all NWs considered in the analysis. Corresponding to the volume fraction of coalesced NWs, a such defined coalescence degree is the relevant quantity when examining data from experimental techniques probing the material volume~\cite{Brandt2014,Fernandez-Garrido2014,Hauswald2014a}. Accordingly, the average equivalent-disk diameter of uncoalesced NWs was determined from a normal distribution fitted to the diameter distribution of all uncoalesced NWs~\cite{Brandt2014}.

The total NW number density was calculated taking into account that coalesced aggregates are composed of several NWs. The number of NWs contained in a coalesced aggregate was estimated dividing the cross-sectional area of the aggregate by the average cross-sectional area of uncoalesced NWs.

Continuous-wave micro-photoluminescence (\textmu-PL) experiments were carried out using a HeCd laser ($\lambda = 325$~nm) with an excitation density below $15$~W/cm$^{2}$. The luminescence was dispersed by a monochromator ($80$~cm focal length, $2400$~lines/mm) and detected by a charge-coupled device detector.

\section{Results and Discussion}

\subsection{Control over the morphology and distribution of GaN NW ensembles}\label{sec1}

\begin{figure}
\centering
\includegraphics[width=0.5\columnwidth]{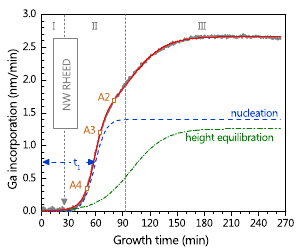}
\caption{Temporal evolution of the Ga incorporation rate per unit area during the growth of sample A1. The time for the appearance of the first GaN-related spots in the RHEED pattern is indicated in the figure. The incubation, nucleation, and elongation stages are labeled as I, II, and III. The dashed red line shows the fit of Equation~(\ref{equation3}) to the experimental data. The blue dashed and the green dashed-dotted lines indicate the contributions of NW nucleation and collective effects, respectively. $t_1$ depicts the average delay time for NW formation, in this case $60$~min. For samples A2--A4, the substrate temperature was increased by $30^{\circ}$C at the indicated times.}
\label{fig:Figure1}
\end{figure} 

\begin{table}[b]	
	\caption{Summary of the growth conditions used for samples A1--A4. The impinging fluxes were kept constant at $\Phi_\textrm{Ga} = (5.5 \pm 0.5)$~nm/min and $\Phi_\textrm{N} = (11.0 \pm 0.5)$~nm/min.}
	\centering
		\begin{tabularx}{\columnwidth}{lXcccc}
				\br
				& 	& sample A1 & sample A2	& sample A3	& sample A4	\\
				\mr
				step 1 	& substrate temperature (\celsius) 	& 815 	& 815 	& 815 	& 815	\\
						& growth time (min)					& 270	& 80	& 65	& 50	\\
				\mr
				step 2 	& substrate temperature (\celsius) 	& -	 	& 845 	& 845 	& 845	\\
						& growth time (min)					& -		& 190	& 205	& 220	\\
				\br
		\end{tabularx}
	\label{tab:growthconditions_ABCD}
\end{table}

\begin{figure*}
\centering
\includegraphics[width=\textwidth]{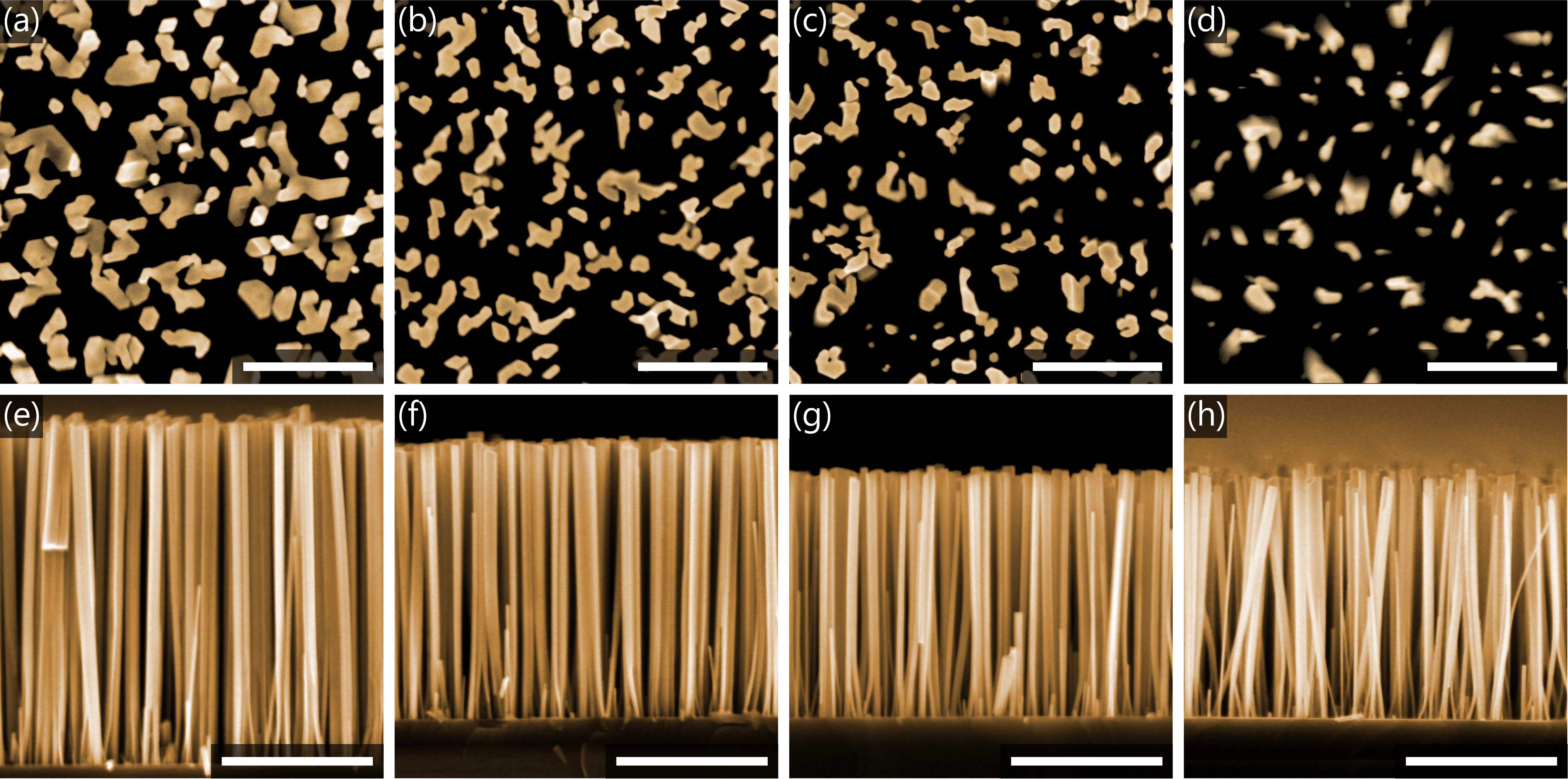}
\caption{Plan-view [(a)--(d)] and cross-section [(e)--(h)] scanning electron micrographs of samples A1--A4, respectively. The scale bars correspond to $500$~nm for plan-view and $1$~\textmu m for cross-sectional micrographs.}
\label{fig:Figure2}
\includegraphics[width=\textwidth]{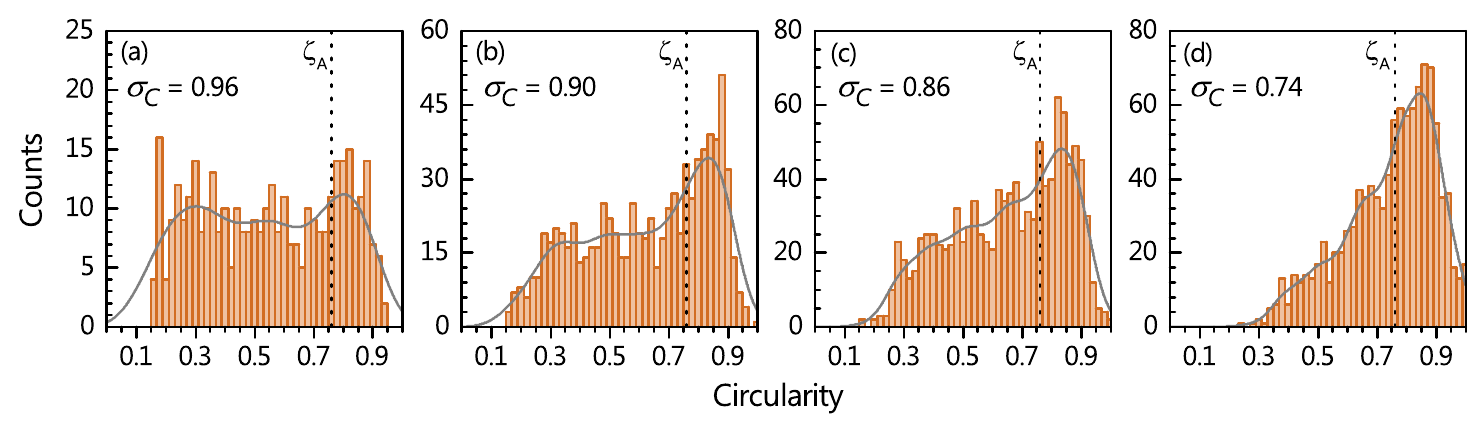}
\caption{ Circularity histograms of the cross-sectional shapes of NWs from samples A1--A4. The solid lines show the kernel density estimation of the respective histogram, the dashed lines indicate the threshold value $\zeta_{A}$. The coalescence degrees $\sigma_{C}$ are derived from eq.~\ref{equation2}.}
\label{fig:Figure3}
\end{figure*}

\begin{figure}
\centering
\includegraphics[width=0.5\columnwidth]{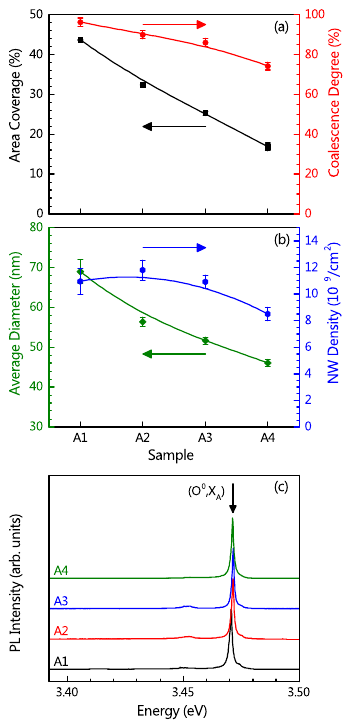}
\caption{ (a) Area coverage, coalescence degree, (b) average uncoalesced nanowire diameter, and total nanowire number density of samples A1--A4 as determined from the statistical analysis of plan-view scanning electron micrographs. The lines are provided as guides to the eye. (c) Normalized photoluminescence spectra at 10 K for samples A1–-A4. The spectra have been shifted vertically.}
\label{fig:Figure4}
\end{figure} 

Here, we investigate to  which extent the morphology and distribution of GaN NWs can be controlled using a two-step growth process where the growth conditions are modified during the nucleation stage. 

Figure~\ref{fig:Figure1} shows the temporal evolution of the Ga incorporation rate per unit area for a reference GaN NW ensemble prepared in a conventional fashion, namely, keeping constant all growth parameters throughout the entire process. This reference sample, referred to as sample A1, was grown at $815^{\circ}$C using a Ga flux of $(5.5 \pm 0.5)$~nm/min. The total growth time was $4.5$~hours. After an incubation time of $25$~min, the appearance of GaN-related spots in the RHEED pattern (not shown here) as well as the initial increase of $\Phi_\textrm{inc}$ (Fig.~\ref{fig:Figure1}) reveal the formation of the first GaN NWs~\cite{Consonni2011,Fernandez-Garrido2015}. Afterward, $\Phi_\textrm{inc}$ rapidly increases due to the continuous nucleation of GaN NWs. As discussed in Ref.~\cite{Fernandez-Garrido2015}, the variation in the slope of $\Phi_\textrm{inc}$ after about $90$~min indicates the end of the nucleation stage. The final increase in $\Phi_\textrm{inc}$ before reaching steady-state growth conditions is due to the onset of collective effects, i.\,e., the shadowing of the impinging fluxes by long NWs and the exchange of Ga atoms between adjacent NWs~\cite{Sabelfeld2013,Fernandez-Garrido2015}. As reported in Ref.~\cite{Fernandez-Garrido2015}, the temporal evolution of $\Phi_\textrm{inc}$ can be described by the sum of two logistic functions,
\begin{equation}\label{equation3}
\Phi_\textrm{inc}=\frac{A_{1}}{1+\exp\left(-\frac{t-t_{1}}{\tau_{1}}\right)}+\frac{A_{2}}{1+\exp\left(-\frac{t-t_{2}}{\tau_{2}}\right)},
\end{equation}
that describe the respective contributions of NW nucleation and collective effects. In Equation~(\ref{equation3}), $t_\textrm{1}$ represents the average delay time for NW formation, $1/\tau_\textrm{1}$ a rate constant related to the NW formation rate after the incubation time, and \textit{A}$_{1}$ the final value of $\Phi_\textrm{inc}$ in the absence of collective effects. Analogously, $t_\textrm{2}$ is the average delay time for the onset of collective effects, $1/\tau_\textrm{2}$ a rate constant, and \textit{A}$_{2}$ the contribution of collective effects to the final value of $\Phi_\textrm{inc}$. As shown in Fig.~\ref{fig:Figure1}, Eq.~(\ref{equation3}) yields an excellent fit of the experimental data. In the figure, we also depict the individual contributions of the two logistic functions. The average delay time for NW formation derived from the fit is $60$~min.

In view of the fact that each NW experiences a different delay time before it is formed~\cite{Fernandez-Garrido2015}, we expect to gain control over both NW number density and coalescence degree by modifying the substrate temperature at the nucleation stage (stage II in Fig.~\ref{fig:Figure1}), as reported by Carnevale et al~\cite{Carnevale2011}. Since the average delay time for NW formation increases exponentially with substrate temperature~\cite{Fernandez-Garrido2015}. raising the temperature before completion of the NW nucleation should suppress further nucleation and thus decrease the final NW number density and coalescence degree. To investigate this possibility, we prepared a series of samples where the substrate temperature was increased by $30^{\circ}$C at different times during the nucleation stage (samples A2--A4). For all samples, we used the same impinging fluxes, initial substrate temperature, and total growth time as for sample A1. As indicated in Fig.~\ref{fig:Figure1}, for samples A2--A4 the substrate temperature was increased after $80$, $65$, and $50$~min, respectively. Therefore, before increasing the temperature, for sample A2 the nucleation was close to the end, for sample A3 well advanced, and for sample A4 at the early times. The growth conditions of samples A1--A4 are summarized in Table~\ref{tab:growthconditions_ABCD}.

Figures~\ref{fig:Figure2}~(a)--(d) show plan-view scanning electron micrographs of samples A1--A4, respectively. The introduction of a two-step growth procedure leads to a clear reduction in the area fraction covered by GaN NWs. Also, the earlier the substrate temperature is raised during nucleation, the stronger the influence on the final NW ensemble. Figures \ref{fig:Figure2} (e)--(h) show the corresponding cross-sectional scanning electron micrographs. The NWs in sample A1 are about $2.2$~\textmu m long. When taking into account the average delay time for the NW formation ($60$~min), we find that the axial growth rate is approximately $10.5$~nm/min. This value is close to the impinging active N-flux, in good agreement with the experiments reported in Refs.~\cite{Yoshizawa1997,Songmuang2007,Tchernycheva2007,Landre_apl_2008,Songmuang2010,Fernandez-Garrido2013a,Schuster2014a,Fernandez-Garrido2015}. In contrast, the average NW length for samples A2--A4 is only $1.6-1.7$~\textmu m. The reduction in the axial growth rate, which becomes Ga-limited, is due to the enhanced Ga desorption during the second step of the growth~\cite{Heying2000}.

Figure~\ref{fig:Figure3} depicts the circularity histograms derived from the analysis of the cross-sectional shapes of the GaN NWs of samples A1--A4. The histogram of the reference sample [Fig.~\ref{fig:Figure3}(a)] is rather broad, reflecting a wide variety of cross-sectional shapes as result of NW coalescence. Interestingly, the variation in the substrate temperature during the nucleation stage has a strong impact on the distribution of cross-sectional shapes. As shown in Figs.~\ref{fig:Figure3}~(b)--(d), the earlier the temperature is raised during the nucleation stage, the narrower the circularity histogram. 

Figure~\ref{fig:Figure4}(a) shows the coalescence degrees for samples A1--A4 derived from their circularity histograms as well as the area fraction covered by GaN NWs. The coalescence degree steadily decreases from $96$ (reference sample) to $74\%$ (sample A4) when decreasing the time at which the substrate temperature is increased. The figure also evidences a decrease in the area fraction from $44$ to $17\%$. In principle, the reduction of both the coalescence degree and the area fraction can be caused by the suppression of further nucleation and/or a decrease in radial growth during the second growth step. 

Next, we extract the average diameter of uncoalesced NWs and the total NW number density. Figure~\ref{fig:Figure4}(b) presents the values of these parameters for samples A1--A4. The average NW diameter steadily decreases from $70$ to $45$~nm from sample A1 to A4. Regarding the total NW number density, the effect of modifying the substrate temperature during growth is not as clear. Increasing the temperature during the second half of the nucleation stage does not seem to influence the total NW number density which remains almost constant ($\approx~1.1 \times 10^{10}$~cm$^{-2}$). Only when the temperature is increased at the beginning of the nucleation stage, we observe a clear reduction to a value of $8.5 \times 10^{9}$~cm$^{-2}$ for sample A4. Therefore, we conclude that the continuous decrease in the coalescence degree observed in Fig.~\ref{fig:Figure4}(a) is mainly caused by a reduction in radial growth during the second step of the growth. This effect is the result of a decrease in the effective Ga/N ratio due to the enhanced Ga desorption.

The high coalescence degree of sample A4 despite exhibiting a low total NW number density as well as a small average NW diameter is surprising. A close inspection of the scanning electron micrograph shown in Fig.~\ref{fig:Figure2}(h) reveals that a significant number of NWs is bent. In addition, the distance between coalesced aggregates is much larger than the average NW diameter [see Fig.~\ref{fig:Figure2}(d)]. We suggest that NW coalescence is not only due to NW mutual misorientation and NW radial growth but also induced by electrostatic attraction during growth~\cite{Brandt2014,Sun2014,Carapezzi2014}. The latter effect is most likely caused by the exposure of the NW ensemble to electrons originating from the N plasma source or the RHEED measurements. Electrostatic attraction between adjacent NWs has already been reported and systematically investigated in Si NWs~\cite{Sun2014}. This phenomenon is expected to be more pronounced for thin and long NWs, as those of sample A4. The intrinsic coalescence degree of sample A4 is therefore expected to be below the measured $74\%$.

Figure~\ref{fig:Figure4}(c) shows normalized photoluminescence spectra at 10 K for samples A1–-A4. The intensities of all samples are comparable despite the reduction in area coverage [Fig. \ref{fig:Figure4}(a)]. For all samples, the spectra are dominated by the recombination of A excitons bound to neutral oxygen [\OoxA] which shows a full width at half maximum of $1.1\pm0.1$~meV. These findings indicate that the given set of growth conditions used affect neither the inhomogeneous strain in the nanowire, nor the density of nonradiative recombination centers.

The results of these experiments demonstrate that, independent of the impinging fluxes, it is possible to obtain a certain degree of control over the morphology and distribution of GaN NWs by increasing the substrate temperature during the nucleation stage. This growth approach is found to be indeed an efficient method to decrease the average NW diameter as well as the area coverage. However, it seems that reduced NW diameters make the NWs more susceptible to electric attraction, leading to NW bending and non-intrinsic NW coalescence. Therefore, despite the reduction in NW diameter, the presented NW ensembles still suffer from a non-negligible degree of coalescence. We anticipate that additional measures aimed to prevent electrostatic attraction of NWs~\cite{Hill2010}, such as negatively biasing the substrate to deflect electrons originating from the N plasma, may enable a further reduction of the NW coalescence. The total NW number density, however, cannot be significantly decreased when growing long and homogeneous NW ensembles. This fact indicates that during the second step of the growth further nucleation is not completely suppressed. Our results are in striking contrast to those obtained by Carnevale et al.~\cite{Carnevale2011}, namely, a high degree of control over NW density by using a similar two-step growth approach. The underlying reason for this discrepancy could be the much shorter growth times used by Carnevale et al. ($<90$~min). The shorter times resulted not only in a lower density but also in much shorter and inhomogeneous NW ensembles. These apparently contradictory results can be reconciled by assuming that despite the presumable unfavorable nucleation conditions the NW number density increases slowly but steadily during the second step of the growth until a homogeneous NW ensemble (as those shown in Fig.~\ref{fig:Figure2}) is formed due to the onset of collective effects~\cite{Sabelfeld2013}.

\subsection{Reduction of the growth time preserving the morphological and optical properties of the NW ensemble}\label{sec2}

\begin{figure}[!t]
\centering
\includegraphics[width=0.5\columnwidth]{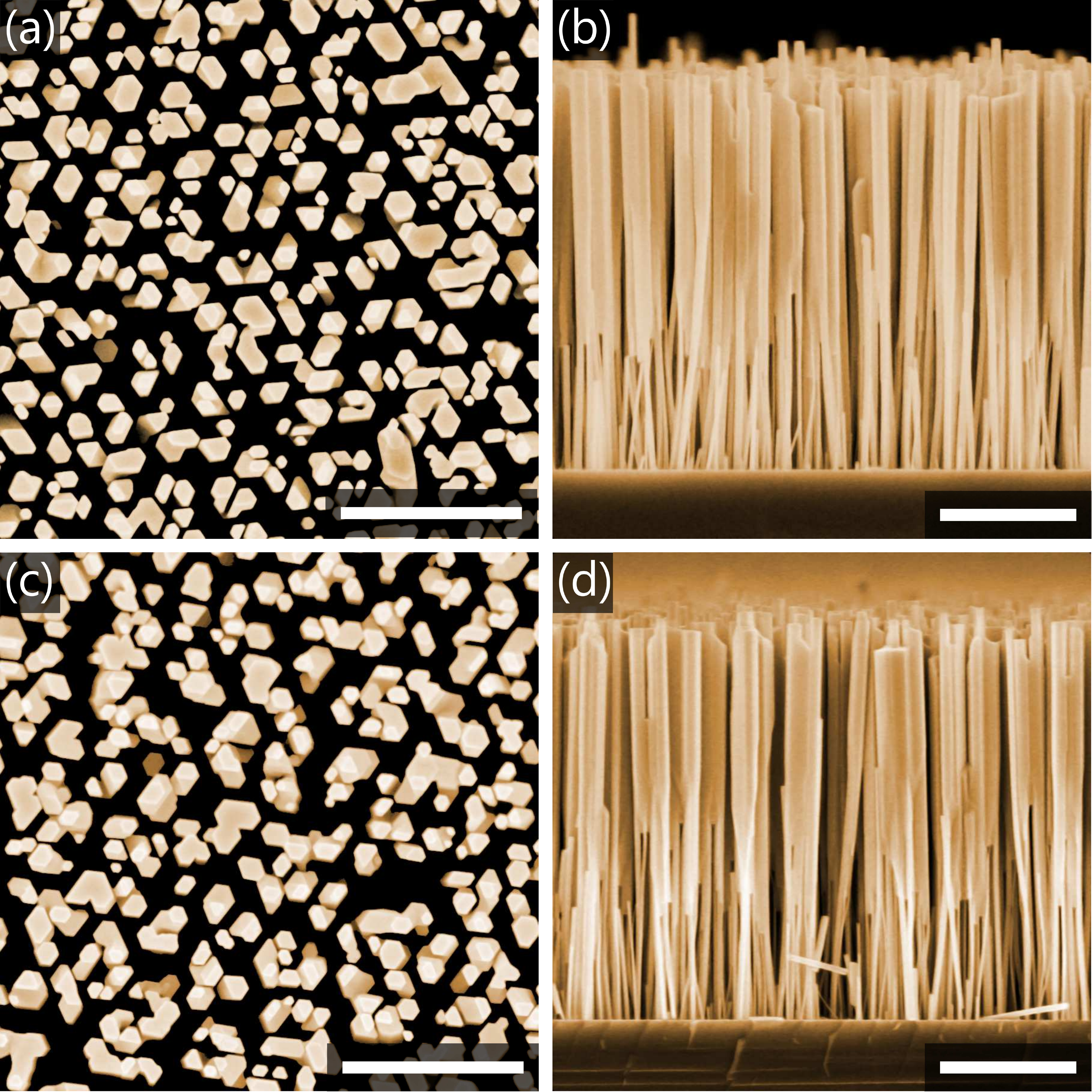}
\caption{ Plan-view [(a) and (c)] and cross-sectional [(b) and (d)] scanning electron micrographs of samples B1 and B2, respectively. The scale bars correspond to $500$~nm for plan-view and $1$~\textmu m for cross-sectional micrographs.}
\label{fig:Figure5}
\includegraphics[width=0.5\columnwidth]{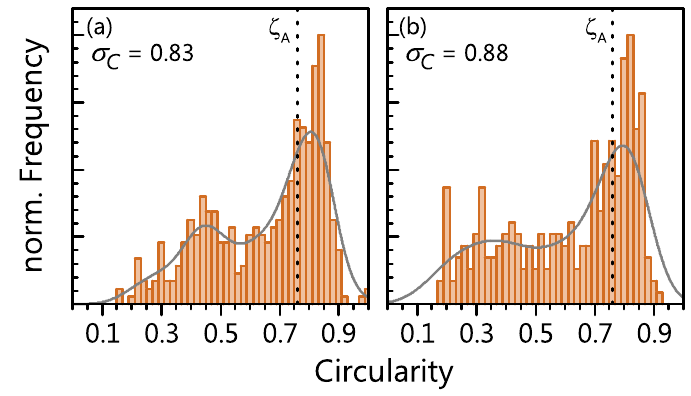}
\caption{Circularity histograms of the cross-sectional shapes of the GaN NWs from (a) samples B1 and (b) sample B2. The solid lines show the kernel density estimation of the respective histogram, the dashed lines indicate the threshold value $\zeta_{A}$. The coalescence degrees $\sigma_{C}$ are derived from eq.~\ref{equation2}.}
\label{fig:Figure6}
\end{figure} 

\begin{table}[b]	
	\caption{Summary of the growth conditions used for samples B1 and B2. The impinging N flux was kept constant at $\Phi_\textrm{N} = (11.0 \pm 0.5)$~nm/min.}
	\centering
		\begin{tabularx}{\columnwidth}{lXcc}
				\br
				& 	& sample B1 & sample B2\\
				\mr
				step 1 	& substrate temperature (\celsius) 		& 855 	& 815	\\
						& growth time (min)						& 420	& 25	\\
						& Ga flux $\Phi_\textrm{Ga}$ (nm/min)	& 16.5	& 5.5	\\
				\mr
				step 2 	& substrate temperature (\celsius) 		& - 	& 855	\\
						& growth time (min)						& -		& 335	\\
						& Ga flux $\Phi_\textrm{Ga}$ (nm/min)	& -		& 16.5	\\
				\br
		\end{tabularx}
	\label{tab:growthconditions_EF}
\end{table}

We have recently reported that a two-step growth approach, as the one described above, can also be used to achieve higher substrate temperatures~\cite{zettler2015}. This possibility arises from the fact that the long incubation time that hinders direct growth at elevated temperatures~\cite{Fernandez-Garrido2015} can be arbitrarily reduced by using a lower substrate temperature during the first step of the growth. Next, we investigate whether the total growth time can be reduced, without modifying the morphological and optical properties of the NW ensemble, using a two-step growth approach. 

To this end, we prepared a second reference NW sample, referred to as sample B1, at a substrate temperature of $855^{\circ}$C. Due to the high substrate temperature, nominally Ga-rich growth conditions [$\Phi_\textrm{Ga}=(16.5 \pm 0.5)$~nm/min] were required to compensate for the high Ga desorption rate~\cite{Fernandez-Garrido2015,zettler2015} .
The total growth time was $7$~h and the delay time before detecting the formation of the first GaN NWs by RHEED (i.\,e., the incubation time) was as long as 90~min. The corresponding average delay time for NW formation $t_1$ was $175$~min. We then prepared another sample using a two-step growth approach (sample B2). During the first step of the growth, we used the growth conditions of sample A1, namely, a Ga flux of $(5.5 \pm 0.5)$~nm/min and a substrate temperature of $815^{\circ}$C. After $25$~min, we observed the onset of NW nucleation by RHEED. At that point, we changed to the growth conditions of sample B1 by first increasing the substrate temperature and subsequently the Ga flux. During the entire process, the Ga and N shutters were kept open. Therefore, during NW nucleation and elongation stages, the growth conditions of samples B1 and B2 were the same. Consequently, $t_1$ was reduced to $115$~min and the growth was finished after a total growth time of $6$~h, i.\,e., $1$~h less than for sample B1. The growth conditions of both samples are summarized in Table~\ref{tab:growthconditions_EF}.

Figures~\ref{fig:Figure5} shows plan-view [(a) and (c)] and cross-sectional [(b) and (d)] scanning electron micrographs of samples B1 and B2. Despite the different growth conditions used during the incubation stage and the shorter total growth time employed for growing sample B2, the sample morphology is indistinguishable. For both samples, the NWs exhibit a similar density as well as comparable lengths and cross-sectional shapes. Interestingly, the cross-sectional scanning electron micrographs reveal that the diameter of these GaN NWs increases during growth. Such a temporal evolution in the NW diameter, not observed before in NW ensembles grown at lower temperatures, suggests that the effective Ga/N ratio is not constant during growth at $855^{\circ}$C for the impinging fluxes used in these experiments~\cite{Fernandez-Garrido2013a}.

Figure~\ref{fig:Figure6} depicts the circularity histograms of the cross-sectional shapes of NWs from samples B1 and B2. Due to the increase in NW diameter during growth, the coalescence degrees were derived from the circularity histograms without introducing a diameter limit for uncoalesced NWs. The coalescence degrees are indicated in Fig.~\ref{fig:Figure6} and listed in Table~\ref{tab:properties_EF}, where we also show the values of the average length, diameter, area coverage, and total number density. As expected from the visual inspection of Fig.~\ref{fig:Figure5}, the quantitative analysis of the scanning electron micrographs reveals that, within the experimental error, the morphological properties of samples B1 and B2 agree fairly well. 

\begin{table}[b]	
	\caption{NW length, area coverage, coalescence degree, diameter and NW number density of samples B1 and B2 as determined from the analysis of scanning electron micrographs.}
	\centering
		\begin{tabularx}{\columnwidth}{X rl c rl}
				\br
								& \multicolumn{2}{c}{sample B1} &\xspace\xspace & \multicolumn{2}{c}{sample B2}			\\
				\mr
				length (\textmu m)							& $3.0$	& $\pm~0.1$	&& $3.0$	& $\pm~0.1$		\\
				area coverage ($\%$)						& $46$	& $\pm~1$	&& $44$	& $\pm~1$		\\
				coalescence degree ($\%$)					& $83$	& $\pm~2$	&& $88$	& $\pm~2$		\\
				average uncoalesced diameter (nm)			& $111$	& $\pm~3$ 	&& $111$	& $\pm~4$		\\
				NW number density ($10^{9}/\textrm{cm}^{2}$)	& $4.2$	& $\pm~0.4$	&& $4.0$	& $\pm~0.4$		\\
				\br
		\end{tabularx}
	\label{tab:properties_EF}
\end{table}

\begin{figure}
\centering
\includegraphics[width=0.5\columnwidth]{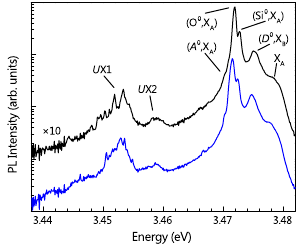}
\caption{ Low-temperature ($10$~K) PL spectra of samples B1 (black, top, multiplied by a factor of 10 for clarity) and B2 (blue, bottom).}
\label{fig:Figure7}
\end{figure} 

Finally, the low-temperature ($10$~K) near band-edge PL spectra of samples B1 and B2 are shown in Fig.~\ref{fig:Figure7}. In both cases, the spectrum is dominated by the recombination of A excitons bound to neutral O [\OoxA] and Si donors [\SioxA] at $3.471$ and $3.472$~eV, respectively. Due to the high substrate temperature~\cite{Corfdir2014,zettler2015}, these transitions are as narrow as $0.7$~meV. Beside these lines, we also observe the recombination of B excitons bound to neutral donors [\doxB], free A excitons [\xA], A excitons bound to neutral acceptors [\aoxA], and the so-called UX band~\cite{Calleja2000,Corfdir2015,Brandt_prb_2010,Pfuller_prb_2010,Sam-Giao2013} [\uxA, \uxB]. For both samples, all these transitions are centered at the same energy and exhibit comparable linewidths and intensities. Therefore, the PL spectra and intensities of samples B1 and B2 are quite similar.

The strong similarities between the morphological and optical properties of samples B1 and B2 reveal that the growth conditions employed during the incubation stage, i.e., prior to NW nucleation, do not influence the properties of the final NW ensemble. By choosing appropriate growth conditions the incubation time can thus be reduced to arbitrary values. Thereby, the present two-step growth approach enables the growth of NW ensembles in shorter times without affecting their final properties.

\section{Summary and conclusions}

We have investigated how a variation in the growth parameters during either the incubation or the nucleation stages influences the final properties of homogeneous GaN NW ensembles prepared by PA-MBE. As a result, we gained both valuable insight into the nucleation mechanisms of spontaneously formed GaN NWs and developed growth methods to improve the control over the spontaneously formed NWs. We demonstrated that in contrast to what would be expected the growth conditions used during the incubation stage influence neither the morphological properties nor the low temperature PL spectra of NW ensembles. Therefore, it is possible to obtain NW ensembles with similar properties but in shorter growth times by using more favorable growth conditions for NW nucleation (lower substrate temperature and/or higher impinging fluxes)~\cite{Fernandez-Garrido2015} during the incubation stage. This finding is important for NW growth at higher substrate temperatures where the incubation time becomes the limiting factor~\cite{Fernandez-Garrido2015,zettler2015}. In contrast, a variation in the growth parameters during the nucleation stage has a strong influence on the properties of the final NW ensemble. The impact on the final morphology depends on the time at which the growth conditions are modified during the nucleation stage. This growth approach does not result in a significant reduction in the NW number density because further nucleation is not completely suppressed after modifying the growth parameters. However, a two-step growth approach is found to be an efficient method to gain control over other important parameters such as area coverage, coalescence degree, and average NW diameter. In order to further reduce the coalescence degree of spontaneously formed GaN NWs, additional measures have to be taken to prevent the electrostatic attraction of thin NWs.

\ack

We thank Anne-Kathrin Bluhm for providing the scanning electron micrographs presented in this work, Vladimir M. Kaganer for fruitful discussions on the attraction and coalescence of NWs, Hans-Peter Schönherr for his dedicated maintenance of the MBE system, and Christian Hauswald for a critical reading of the manuscript. Financial support of this work by the Deutsche Forschungsgemeinschaft within SFB 951 is gratefully acknowledged.


\end{document}